# Usage Frequency and Application Variety of Research Methods in Library and Information Science: Continuous investigation from 1991 to 2021

Chengzhi Zhang[1,*], Liang Tian[1] and Heating Chu[2]

1. Department of Information Management, Nanjing University of Science and Technology, Nanjing, Jiangsu, China

2. Palmer School of Library and Information Science, Long Island University, NY 11548, USA

**Abstract:** The present study analyzed over 26,000 research articles published between 1991 and 2021 in twenty-one major LIS (Library and Information Science) journals, using the machine learning (ML) approach to categorize the research methods used by LIS scholars. The findings of this study are significant. Firstly, there has been a shift in the research strategy from conceptual research (e.g., "Theoretical approach") to empirical research (e.g., "Interview") in LIS investigations over the past 31 years. Secondly, the research topics explored by LIS scholars during this period have moved from system-centered issues (e.g., "Information retrieval/models and algorithms") to user-centered topics (e.g., "Information services "). Thirdly, the study revealed dynamic and revealing relationships between the 18 research topics identified in the study and the 16 research methods commonly adopted in the LIS field. These dynamic relationships can be visualized by year and longitudinally via an interactive map created in this study.

[*] Corresponding author: Chengzhi Zhang (zhangcz@njust.edu.cn).

## 1. Introduction

During the 1970s and 1980s, researchers in the Library and Information Science (LIS) field engaged in significant discussion regarding the future of LIS as an independent discipline (e.g., Harmon, 1971; Schrader, 1984). This discussion continued for several decades and is ongoing today (e.g., Ma, 2012; Sheble, 2016; Steinerová, 2003; Wang, Yang & Wu, 2021). These self-analyses offer valuable insights into the evolution of LIS and provide a clearer understanding of the field's current state and future development. This discussion remains important for the continued advancement of the LIS field today.

Library and Information Science (LIS) claims to be a specific field of study, with its own procedures and methodology, according to LIS epistemology (Risso, 2016). Early research in LIS was descriptive in nature, but as the field has evolved, research methods applied in LIS publications have become more diverse and standardized. The highly interdisciplinary nature of LIS means that research methods used in the field are often influenced by other disciplines such as psychology, as well as technological developments (Cano, 1999; Prebor, 2007, 2010). The rapid advancements in technologies such as social media, big data, and machine learning have brought about dramatic changes in LIS research and may alter some of the fundamental concepts of LIS research (Ma & Lund, 2021).

Scholars have been studying research methods in the Library and Information Science (LIS) field for several decades, as evident from research by Peritz (1980) and Schlatcher and Thomison (1974). One of the most influential studies was conducted by Järvelin and Vakkari (1990), who proposed a classification scheme of research methods, comprising five research strategies and seven data collection methods. Many subsequent studies have categorized research methods used in LIS based on Järvelin and Vakkari's (1990) model. Examples of such studies include those conducted by Bernhard and Lambert (1993), Hider and Pymm (2008), and Tuomaala,

Järvelin, and Vakkari (2014). Chu (2015) takes a different approach and creates a taxonomy of research methods based solely on the data collection technique used.

The basic method for conducting research in the field of library and information science (LIS) is manual content analysis. However, a major issue that these studies must confront is the limited number of publications that can be analyzed. With the advancement of machine learning (ML), there are now techniques available for the automatic classification of entities, such as images (e.g., Karpathy & Li, 2015) and queries in question answering (QA) efforts (e.g., Soares & Parreiras, 2020). Our study aims to explore the possibility of using machine learning methods to automatically classify research methods reported in LIS publications between 1991 and 2021.

This study investigates the research methods used in the field of information science (LIS) from two perspectives. Firstly, we examine the frequency of usage and variety of application of research methods. Secondly, we explore the relationship between research topics and research methods. Using a machine learning (ML) approach and a full dataset covering 31 years, we are able to observe changes in the selection and application of research methods in LIS over time. This is an improvement compared to previous studies, which often relied on a single year of data (e.g., Järvelin & Vakkari, 1990) or selected years from a specific period (e.g., Tuomaala, Järvelin & Vakkari, 2014; Yontar & Yalvaç, 2000).

This study employs a machine learning approach to answer the following three research questions (RQs) based on data collected from three major information science (LIS) journals published between 1991 and 2021:

***RQ1:*** Can we obtain a similar level of classification accuracy as the manual approach by using the machine learning approach?

***RQ2:*** What changes have occurred in the usage frequency and application variety of research methods reported in LIS publications between 1990 and 2021?

***RQ3:*** What is the relationship between research methods and research topics in the LIS field during the period of 1990-2021?

## 2. Literature review

### 2.1 Research methods utilized in LIS

The research methods used in the field of information science (LIS) have been explored by researchers through content analysis of research publications (e.g. Chu, 2015; Järvelin & Vakkari, 1990) or through conceptual discussions of research methods applied in LIS (e.g. Ma, 2012; Weller & Haider, 2007). Similar to Chu (2015) and Järvelin and Vakkari (1990), the present study aims to address the three research questions outlined earlier by performing a content analysis of publications from three major LIS journals between 1991 and 2021. However, in contrast to previous studies, this research employs a machine learning (ML) approach instead of relying solely on manual analysis. Through the use of ML, we hope to provide a more comprehensive and accurate analysis of research methods in the LIS field.

Previous studies on research methods in the field of information science (LIS) mainly explored the application of one or a few research methods (Schlatcher & Thomison, 1974; Wallace, 1985). Later, researchers began to address the same research question more comprehensively by developing a categorization scheme of research methods. One of the earliest such schemes was created by Järvelin and Vakkari (1990), which contains 13 research methods and additional categories for "other methods", "no method", and the like. The 13 specific methods are further divided into two kinds: "research strategies" and "data collection methods" (see Table 1). This categorization scheme has been widely adopted and utilized in related research since its creation (Lou et al., 2021; Zhang et al., 2023; Zhang and Tian, 2023). For instance, Hider and Pymm (2008) sed this scheme to study the evolution and distribution of research methods based on a content analysis of high-profile LIS journal literature published in 2005. Table 1 summarizes relevant studies that have utilized this categorization scheme.

| **Authors** | **Source Data** | **Method Classification Scheme** |
| --- | --- | --- |
| Eaton and Burgin (1983) | 250 articles from 62 core journals in 1983 | Theoretical/ analytical research, information system design, surveys on the library public, surveys or experiments on libraries, bibliometric studies, content analysis, secondary analysis, historical method, descriptive bibliography, comparative study, and other methods. |
| Järvelin and Vakkari (1990) | 833 articles from 37 core journals in 1985 | Research Strategies: empirical research strategy, conceptual research strategy, mathematical or logical method, system and software analysis and design, literature review, discussion paper, bibliographic method, other methods, not applicable, no method<br>Data Collection Methods: questionnaire, interview, observation, thinking aloud, content analysis, citation analysis, historical source analysis, several methods of collecting, use of data collected earlier, other methods of collecting, not applicable. |
| Kumpulainen (1991) | 632 articles from 30 core journals in 1975 | The same classification scheme as Järvelin and Vakkari. |
| Järvelin and Vakkari (1993) | 142/359/449 articles from 40 core journals in 1965, 1975, and 1985 respectively | Same as Järvelin and Vakkari (1990) |
| Blake (1994) | LIS doctoral dissertations in 1975-1989 | Descriptive, case study, bibliometric, historical, questionnaire/interview, content analysis, modeling, experiment, theoretical approach, and mixed methods. |
| Cano (1999) | 354 articles from two core Spanish journals in 1977-1994. | Same as Järvelin and Vakkari (1990) |
| Yontar and Yalvaç (2000) | 817 articles in 1952-1994 | Same as Järvelin and Vakkari (1990) |
| Hider and Pymm (2008) | 567 articles from 20 core journals in 2005 | Same as Järvelin and Vakkari (1990) |
| Tuomaala, Järvelin and Vakkari (2014) | 142/449/626 articles from core journals in 1965, 1985, and 2005 respectively | Same as Järvelin and Vakkari (1990) |
| Chu (2015) | 1,162 articles published by JDoc, JASIST, and LISR from 2001-2010 | Bibliometrics, content analysis, Delphi study, ethnography/field study, experiment, focus groups, historical method, interview, observation, questionnaire, |
| | | research diary/journal, theoretical approach, think-aloud protocol, transaction log analysis, webometrics, and other methods. |
| Chu and Ke (2017) | 1,981 articles published by JDoc, JASIST, and LISR from 2001-2010 | A classification scheme that is the same as Chu's (2015). |
| Zhang, Tam, and Cox (2021) | 2,599 articles from JDoc, JASIST, and LISR from 2008-2018 | Based on the scheme by Chu (2015) and expanded to 29 research methods: Bibliometrics, content analysis, Delphi study ethnography/field study, experiment, focus groups, historical method, interview, observation, questionnaire, research diary/journal, theoretical approach, think aloud protocol, transaction log analysis, webometrics, agent-based modeling/ simulation, annotation, classification, clustering, comparative evaluation, document analysis, information extraction, IR related indexing/ranking/ query methods, mixed method, digital ethnography, network analysis, statistical methods, topic modeling, other methods. |
| Ma and Lund (2021) | 3,422 articles from six core journals in 2006, 2012 and 2018 | A scheme similar to Tuomaala, Järvelin and Vakkari (2014). |
| Lund and Wang (2021) | 6,000 articles between 1980 to 2019 in five top, practitioner-oriented LIS journals | A scheme similar to Tuomaala, Järvelin and Vakkari (2014). |
| Lou et al. (2021) | 2,421 articles by 53 Chinese scholars in LIS field between 1980 and 2020 | A classification scheme similar to Chu (2015). |
| Zhang et al. (2023) | 5,281 articles between1990 to 2019 from JDoc, JASIST, and LISR | A classification scheme that is the same as Chu's (2015). |
| Vakkari, Järvelin and Chang (2023) | 631/711/1,210 articles from core journals in 1995, 2005, and 2015 respectively | A scheme similar to Tuomaala, Järvelin and Vakkari (2014). |

**Table 1. Summary of related studies on research methods applied in LIS**

To ensure that all categorized research methods are collectively exhaustive and mutually exclusive, Chu (2015) used a single criterion, data collection technique, instead of data analysis technique or a combination of both, to classify research methods. Based on this approach, Chu (2015) developed a scheme of 16 categories of research methods reported in LIS publications between 2001 and 2010. Building upon Chu's framework, Zhang,

Tam, and Cox (2021) expanded the total number of research methods to 29 to incorporate methods used in articles on AI and big data, such as classification and clustering.

The studies summarized in Table 1 examine the research methods used in LIS publications and their frequency through empirical analyses. These studies rely on manual content analysis to identify and classify the research methods reported in the publications. However, the manual approach is time-consuming, which limits the number of publications that can be analyzed. As a result, most studies on research methods in LIS analyze only a limited number of publications for a given year (e.g., Järvelin & Vakkari, 1990) or selected years over a period (e.g., Ma & Lund, 2021). To overcome this limitation, this study adopts the machine learning approach to classify research methods in a larger dataset covering three major LIS journals over a period of 31 years (1991-2021). This approach enables us to observe changes in research method selection and application and explore the relationships between research methods and research topics in the LIS field.

**2.2 The relationship between research topics and research methods in LIS**

Research methods are closely linked to the topics that scholars intend to study. Thus, choosing the right research method based on what is being explored helps to better carry out the study (Lowhorn, 2007). In the field of LIS, the application of research methods is often examined in light of research topics investigated.

Järvelin and Vakkari (1990) revealed the relationship between research methods and research topics by doing a cross-tabulation of the two. Tuomaala, Järvelin, and Vakkari (2014) extended this research and examined changes in research topics and research methods in LIS at three different time points: 1965, 1985, and 2005. Their findings show that research on library and information services decreased after 1985, while research on information-seeking and scientific communication grew during the same period. The authors also found that the proportion of empirical research strategies was high and increased over time, with the survey method being the most frequently used method.

In recent years, researchers have conducted studies to investigate the relationship between research methods and research topics in the field of Library and Information Science (LIS). Järvelin and Vakkari (1990) and Tuomaala, Järvelin, and Vakkari (2014) examined changes in research topics and research methods in LIS over time. Ma and Lund (2021) also performed a content analysis to study the evolution and distribution of LIS research topics and data collection methods. They found that social media and data science topics played a larger role in LIS research in 2018 than in previous years.

VanScoy and Fontana (2016) analyzed articles on references and information services (RIS) and found that quantitative approaches were more commonly used than qualitative approaches. They suggested that a greater diversity of approaches and methods could benefit the study of RIS. Gore et al. (2009) examined research articles in health science librarianship and found that a variety of research methods and techniques were being employed, with a focus on the behavior, attitudes, and opinions of library users.

While most of the studies in this area have used a manual content analysis approach, the present research will adopt Chu's (2015) categorization scheme and use machine learning to investigate the use of research methods in LIS. By doing so, the research aims to provide a comprehensive and up-to-date understanding of research methods applied in the field.

### 2.3 Automatic classification of research method in academic articles

The literature reviewed up to this point shows that most relevant studies have used the manual approach, which limits researchers from handling a large amount of collected data. With the advancement of machine learning technology, some scholars are now attempting automatic classification of research methods. Automatic classification of research methods refers to the automatic determination of research methods used in an article. Text classification, which is one of the most fundamental tasks in natural language processing (NLP), has a wide range of applications, such as sentiment recognition (e.g. Aghajanyan et al., 2021; H. Liu et al., 2021; Muñoz & Iglesias, 2022) and question answering (e.g., Arora et al., 2020; Mehri & Eric, 2021). Although research method classification and sentiment recognition are both text classification tasks, research method classification is more challenging. This is because the classification of research methods involves more specialized knowledge. Moreover, information on the research method only accounts for a small portion of the full text, which may lead to a lot of noise when classifying the text.

A scholarly article may report an investigation that uses multiple research methods. Efforts in automatic identification and classification of research methods in these types of studies are multi-label text classification tasks. Traditional multi-label classification methods mainly consist of two kinds: problem transformation and algorithm adaptation (Zhang & Zhou, 2014). Using deep learning techniques, multi-label classification involves finding the features through deep neural networks while employing a multi-label classifier on top (W. Liu et al., 2021). Many different deep learning techniques are applied in multi-label classification (e.g. Chalkidis et al., 2019; R. Wang et al., 2021). A detailed comparison of these deep learning techniques is beyond the scope of this study and will be covered in a separate article.

Previous studies have attempted the task of automatic classification of research methods. For instance, using traditional multi-label classification methods, Eckle-Kohler, Nghiem, and Gurevych (2013) experimented with the automatic classification of research methods reported in the domain of social sciences and achieved 0.68 as their highest classification accuracy. Zhang, Tam, and Cox (2021) chose an automated, rule-based method to identify and classify research methods reported in LIS scholarly articles in addition to content analysis and text mining methods. They suggested the need for a new approach to classifying LIS research methods to capture the complex structure and relationships between different aspects of methods. Rule-based methods rely on the occurrence of feature words to determine whether a method is used in an article. However, when an article mentions a method, it does not necessarily use that method. It might well be that, for instance, a method is mentioned for comparison purposes. Zhang, Li, and Chu (2020) also explored the automatic classification of research methods in the LIS domain using traditional machine learning methods, but the classification accuracy was low. Thus, the present study chooses deep learning techniques instead of the traditional ones to classify research methods reported in LIS publications.

## 3. Methodology

This study analyzes changes in research method usage in the LIS field and examines the relationship between research topics and research methods over a 31-year period. It comprises three main components: automatic classification of research methods (Section 3.2), topic modeling (Section 3.3), and data analysis (Section 3.4). The dataset of LIS research articles from Chu and Ke (2017) was expanded from 2001-2009 to 1991-2021. A detailed methodology for this study is presented below.

### 3.1 Dataset

Twenty-one journals in LIS were selected for this study, which have also been examined in previous studies (e.g. Vakkari, Chang, & Järvelin, 2022; Fidel, 2008). We collected metadata of research articles published from 1990 to 2021 in these twenty-one journals, and Table 2 shows the number of articles collected from each journal. Out of the total number of articles, 1977 were already annotated using the categorization scheme of research methods by Chu and Ke (2017), and this annotated dataset constitutes our preliminary training data.

| Journal | # of Articles |
|---|---|
| *Aslib journal of information management* | 998 |
| *College & research libraries* | 1,077 |
| *Electronic library* | 1,298 |
| *Information & culture* | 526 |
| *Information processing & management* | 2,265 |
| *Information research-an international electronic journal* | 739 |
| *Information technology and libraries* | 600 |
| *International journal of information management* | 1,713 |
| *Journal of documentation* | 1,189 |
| *Journal of information science* | 1,296 |
| *Journal of librarianship and information science* | 737 |
| *Journal of the association for information science and technology* | 3,837 |
| *Library & information science research* | 732 |
| *Library collections acquisitions & technical services* | 501 |
| *Library quarterly* | 452 |
| *Library resources & technical services* | 480 |
| *Library trends* | 1,164 |
| *Libri-international journal of libraries and information studies* | 709 |
| *Online information review* | 1,337 |
| *Reference & user services quarterly* | 357 |
| *Scientometrics* | 4,958 |
| **Total** | **26,965** |

**Table 2. Number of articles from each journal**

### 3.2 Coding scheme and classification of research methods

To identify and classify research methods in articles, this study utilizes coding scheme of Chu and Ke (2017). This section aims to complete the classification of research methods for 26,965 academic articles. Figure 1 illustrates the process of implementing the method classification. Out of the 26,965 academic articles, 1,977 were manually annotated by Chu and Ke (2017). For the remaining 24,988 articles, this study attempts to use a machine learning approach to achieve automatic classification.

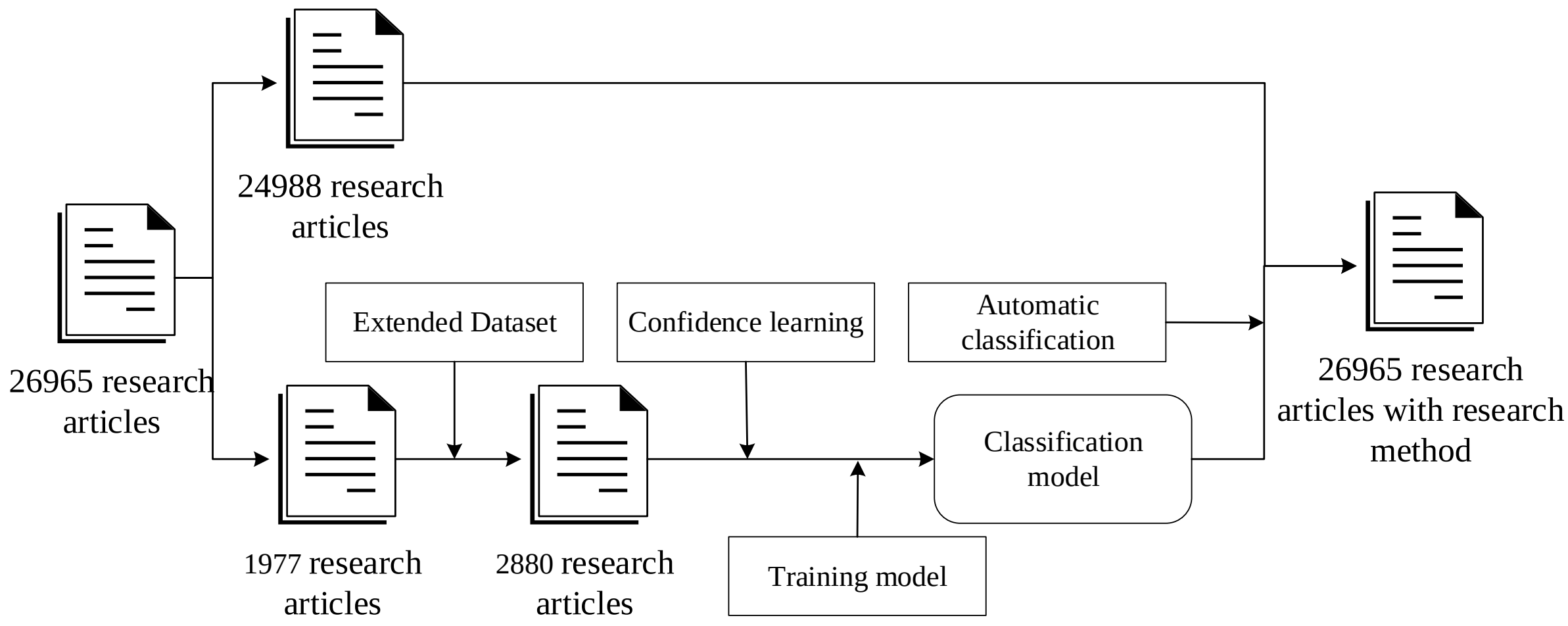

**Figure 1. Automatic classification of research methods**

### 3.2.1 Training dataset

The dataset used in our study consists of 26,965 research articles published in 21 core journals in the LIS field from 1991 to 2021. It should be noted that not all abstracts of research papers explicitly mention the specific research methods utilized. In a study by Zhang et al. (2020), a small-scale experimental comparison was conducted to assess the classification of research methods based on abstracts versus full-text analysis. Upon analyzing the results, the researchers found that while the abstract-based method classification (Eckle-Kohler et al., 2013) was somewhat weaker than the full-text-based approach (Zhang et al., 2020), the difference was deemed acceptable. Moreover, obtaining the full text of all 26,965 research articles in the Library and Information Science (LIS) field from 1991 to 2021 presented significant challenges. Therefore, as a compromise solution, we classify research methods based on the abstract of the article.

We chose to use the research method categorization scheme developed by Chu and Ke (2017) because it adheres to the two fundamental principles in classification: collective exhaustiveness and mutual exclusiveness of all the categories contained within. The schema of Chu and Ke (2017) consists of 16 methods, namely bibliometrics, content analysis, Delphi study, ethnography/field study, experiment, focus groups, historical method, interview, observation, questionnaire, research diary/journal, theoretical approach, think aloud protocol, transaction log analysis, webometrics, and other methods. For detailed definitions of these research methods, please refer to Appendix A.

The frequency distribution of research methods reported in the dataset from Chu and Ke (2017) is shown in the blue bars of Figure 2 after removing articles without abstract. The intercoder agreement rate of the data set was 86.7%. As seen, some categories (e.g., Delphi study) have too few cases in the dataset. Extreme sample imbalance can severely affect machine learning performance, so a technique was devised to expand each method category with less than 300 articles in the dataset. The number 300 was chosen because a lower frequency was found to be less effective in automatic classification in a pilot study.

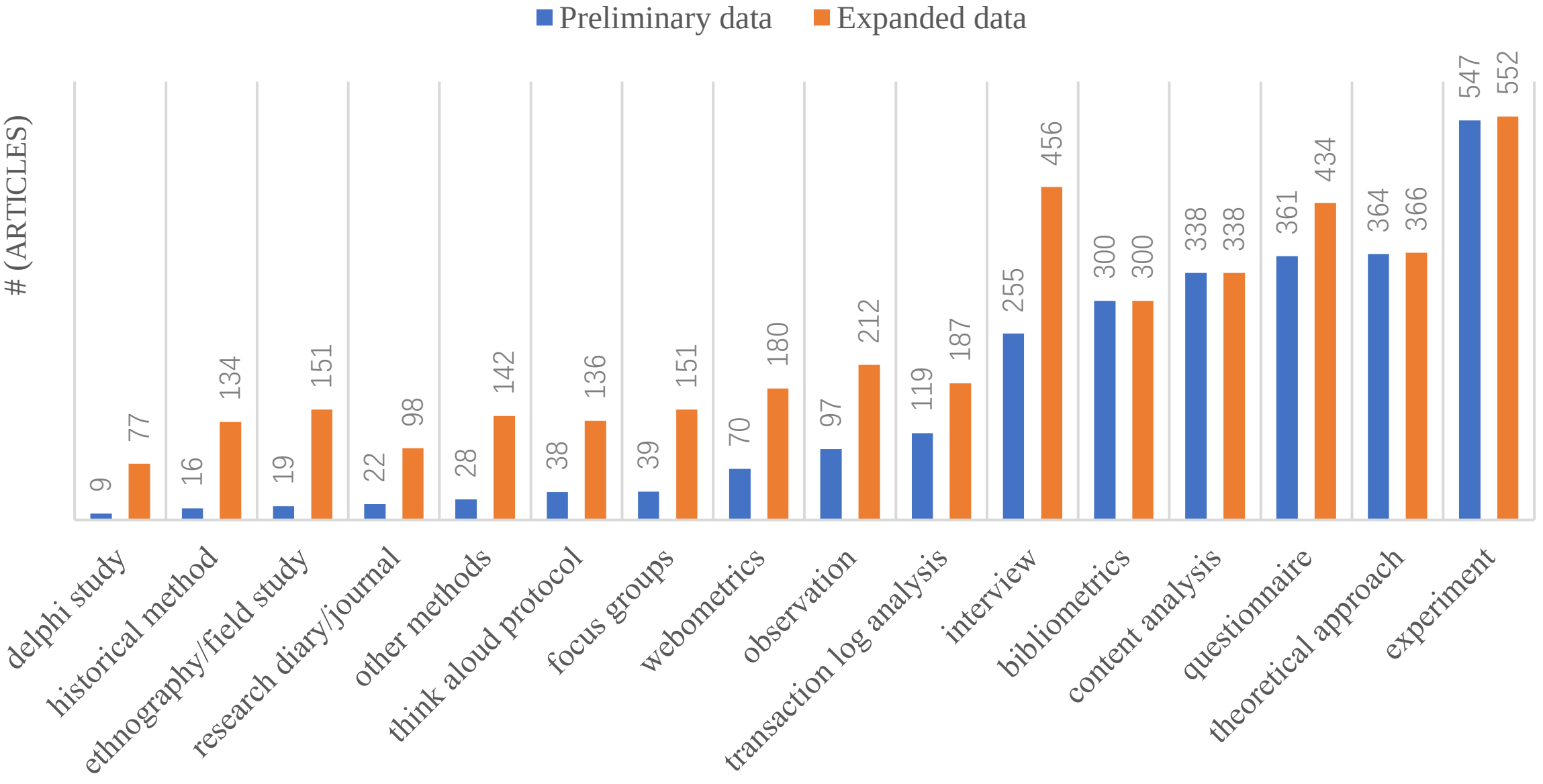

**Figure 2. Distribution of original and expanded datasets by research method**

The technique for expanding those method categories with less than 300 articles consists of four steps:

Firstly, TextRank is used to extract 20 keywords for each article in the dataset from Chu and Ke (2017) (Mihalcea & Tarau, 2004).

Secondly, the ten most frequently appearing keywords are selected for each category of low-frequency method, as detailed in Appendix B.

Thirdly, Web of Science (http://webofscience.com/) is searched using the chosen keywords for each category, and the abstracts of the 100 most relevant results are downloaded[1]. Because some papers do not directly mention the method name, the results obtained by searching with the method name are limited. This article uses keywords to expand the search range, these keywords may be tools used in the method, or keywords involved in the process of using research methods. For example, the corresponding keywords for using Webometrics are "web" and "measure" (see Appendix B).

Lastly, the retrieval results are manually filtered and annotated, and the labeled data is added to the dataset from Chu and Ke (2017). The ratio of valid documents to the total number of search results is presented in the third column of Appendix B.

Consequently, an expanded dataset (shown in the orange bars in Figure 2) was obtained, which became the new training dataset for the current research. Because one study may use more than one research method, expanding those method categories with less than 300 articles also increased the number of articles in other method categories (e.g., experiment, questionnaire), as shown in Figure 2.

### 3.2.2 Automatic classification of research methods

(1) Classification models

Several models for classification, including SciBert (Beltagy et al., 2019), GRU (Cho et al., 2014) and TextCNN (Zhang & Wallace, 2016), are examined in this research and their details are described below. We also normalize the words in the labelled training data before using Chi-square of bigram (Jiang & Zhai, 2007) for feature selection.

---

[1] Here, we retrieve 100 results (or all if less than 100), mainly to reduce the workload of subsequent manual annotation and ensure the quality of the returned results as much as possible.

As noted earlier, problem transformation and algorithm adaptation are the two basic strategies for multi-label classification (Zhang & Zhou, 2014), a task this study must undertake categorizing research methods. The problem transformation strategy comprises Binary Relevance (BR), Classifier Chain (CC), Label Powerset (LP), and Random k-label sets (RAKEL). We use XGBoost (a scalable tree boosting system) as the basic classification algorithm in our study (Chen & Guestrin, 2016). As for the algorithm adaptation strategy, we choose KNN (k-nearest neighbors) as the basic classifier. A combination of five classification algorithms (i.e., BR-XGB, CC-XGB, LP-XGB, RAKEL-XGB, MLKNN) are selected as the baseline model (Keller, Gray & Givens, 1985). Feature vectors are represented by TF-IDF (term frequency – inverse document frequency) scores of bigram (Salton & Buckley, 1988).

The present study considered several deep learning models before choosing one. SciBert is a Bert model trained on scientific text (Devlin et al., 2018). We input our text data into SciBert for text representation and then use the following natural networks (i.e., GRU and TextCNN) to select features. The first model directly inputs the hidden layer of SciBert to the output layer (see Figure 3a). With the second model, the output of SciBert is inputted into GRU while each document is represented as sequence and hidden units. After passing through the attention layer, an embedding document is computed through the content-aware embedding algorithm with attention scores. Prediction results are obtained through the output layer (see Figure 3b). In the third model, the output of SciBert is inputted into TextCNN. The output of TextCNN is inputted into the output layer to get prediction results (see Figure 3c).

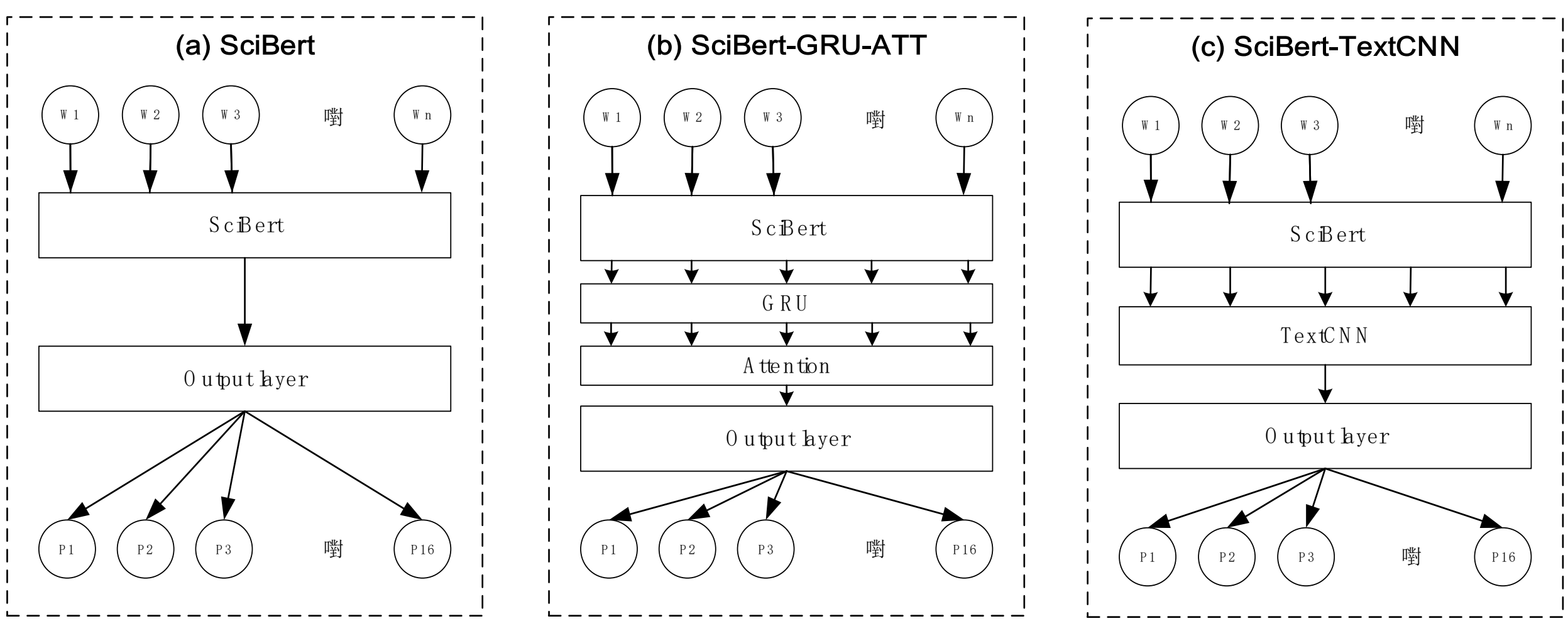

**Figure 3. Illustration of three deep learning models used in this study**

The important parameters of the deep learning models in this study are reflected in the optimizer and loss function. 1) This research uses two kinds of optimizers. The first one is Adam which adaptively sets the learning rate for each floating-point parameter, taking into account the past gradient history ( Amari & Ba, 2017). The other optimizer is stochastic gradient descent --SGD (Amari, 1993). We choose Adam to quickly approach the target, and then use SGD for further optimization. 2) Binary cross entropy (BCE) is used for research method classification as loss function because BCE can handle multi-label classification (Hsieh et al. 2018).

In the three models described above, the best model is chosen for research method classification through comparison and five-fold validation for machine learning to categorize research methods in the unlabeled data.

(2) Confident learning to clean classification dataset

In the process of expanding the dataset, this study may have introduced some mislabeled or noisy data. We decide to use confident learning to find and remove those noisy data. Confident learning is an alternative approach which focuses on label quality by characterizing and identifying label errors in datasets. Based on the principle of noise data pruning, confident learning evaluates the noise by counting and sorts the examples for confident training (Northcutt, Jiang & Chuang , 2021). Confident learning estimates the joint distribution of noisy labels and real labels to fully describe label noise under category conditions. The noisy data with label errors are found and pruned.

We used Cleanlab (https://github.com/cgnorthcutt/cleanlab) to estimate the noisy labels in our dataset. Cleanlab, an open-source tool, is used to clean noisy labels and finding mislabeled data in datasets based on confident learning. The following illustrates what we do in this procedure. 1) We use the expanded dataset containing noise to train all models and identify the best one which is selected for five-fold validation to predict a result for each sample. 2) We choose Cleanlab to identify the noisy data in our dataset. 3) The dataset after noise removal is used for training our final classification model.

(3) Evaluation measures of automatic classification

For testing the classification model, five-fold cross-validation was adopted, while sample-based evaluation measures and label-based evaluation measures were chosen as the evaluation metrics for the classification model (Madjarov et al., 2012). Sample-based evaluation measures comprise of Samples_P, Samples_R, and Samples-$F_1$, and they are calculated based on the P (Precision), R (Recall), and $F_1$ (harmonic mean of precision and recall) for each sample and the mean of all samples. On the other hand, label-based evaluation measures comprise of Micro_P, Micro_R, Micro-$F_1$, Macro_P, Macro_R, and Macro-$F_1$, which require calculating the P, R, and $F_1$ for each label and then calculating the mean of all labels. Equations (1), (2), and (3) represent the base formula for P, R, and $F_1$, respectively. "True positives" represents the correctly predicted quantity, "total predicted positives" represents the quantity predicted to be used, and "total actual positives" represents the used quantity. For automatic classification, we used Micro-$F_1$, Macro-$F_1$, and Samples-$F_1$ as our evaluation metrics.

$$P = \frac{\#\ true\ positives}{\#\ total\ predicted\ positives} \quad (1)$$

$$R = \frac{\#\ true\ positives}{\#\ total\ actual\ positives} \quad (2)$$

$$F_1 = \frac{2PR}{P+R} \quad (3)$$

### 3.3 Research topic modelling

In this study, we employed Top2Vec for topic modeling, which is a popular approach to distributed representations for topic modeling (Angelov, 2020). To evaluate the effectiveness of the topic modeling, we used CV (context-vector-based coherence) as the evaluation metric (Röder, Both & Hinneburg, 2015). A higher CV value indicates better topic consistency and improved topic modeling. Specifically, we calculated CV values for the topic terms obtained after topic modeling and compared the modeling effects with different numbers of topics. As illustrated in Figure 4, we found that Top2Vec provides the best modeling results when the number of research topics is 18.

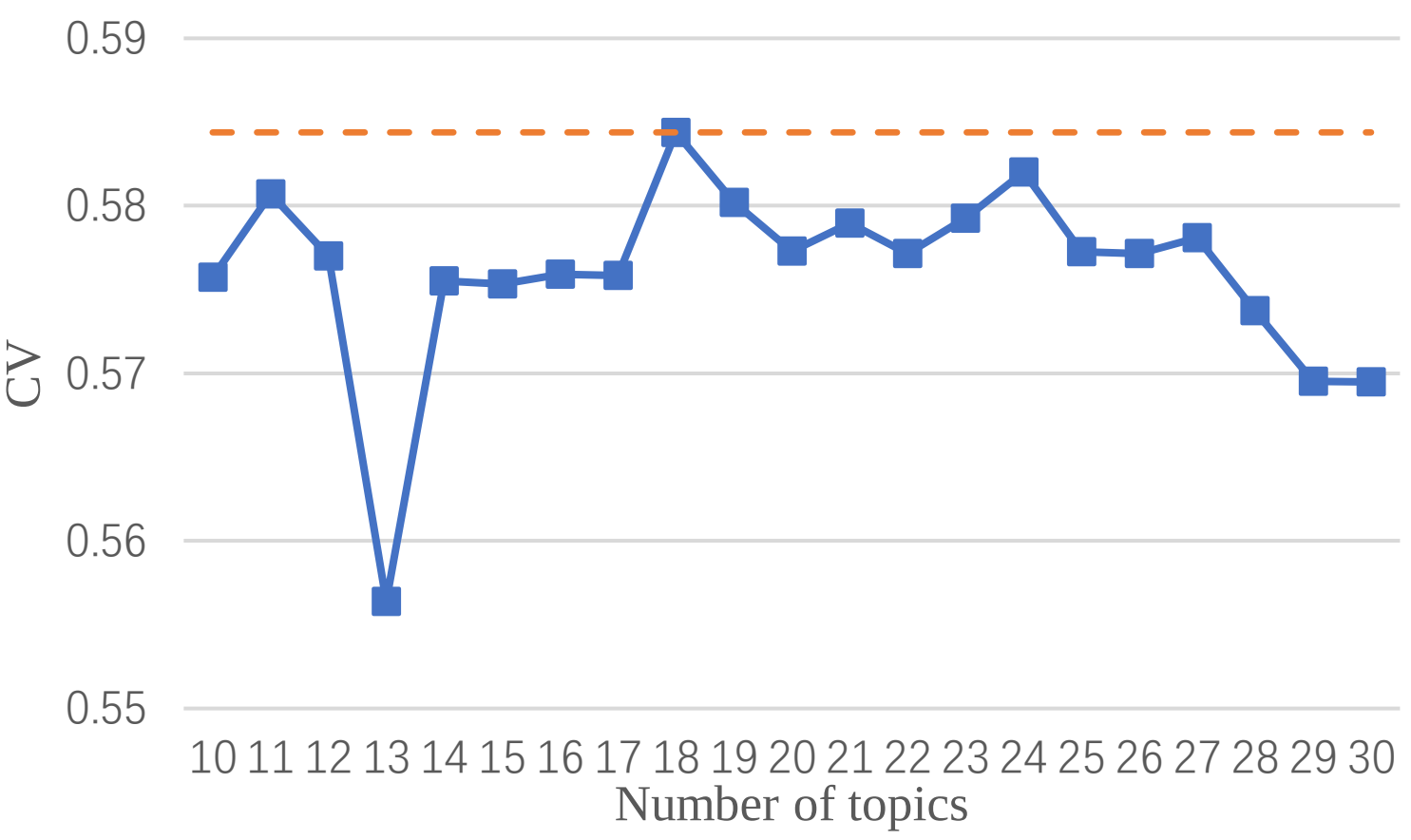

**Figure 4. CV of different topic modeling techniques and a different number of topics.**

As illustrated in Figure 4, we identified 18 research topics using automatic methods. Appendix C presents these topics in alphabetical order. The topics were derived by analyzing and combining specific terms from our dataset using topic modeling. Additionally, we consulted relevant existing research in the field of LIS, including the classification of Järvelin and Vakkari (1993), Tuomaala et al. (2014), Vakkari, et al. (2022), and Vakkari et al. (2023).

### 3.4 Usage frequency and application variety

This study aims to examine the usage of LIS research methods from two perspectives: usage frequency and application variety (i.e., applying one research method for different research topics). Usage frequency is defined as the proportion of articles reporting the usage of different methods in a given year. Equation (4) is the frequency of usage formula of the *i*-th method in j-th year. $N_{i,j}$ is the number of articles reporting the usage of the *i*-th method in the *j*-th year. $\sum_{i=1}^{M} N_i$ is the number of articles published in the *j*-th year. *M* is equal to 16, the total number of research methods examined in this study. We take the average usage frequency of the *i*-th method in all years as the overall usage frequency of the *i*-th method, as shown in Equation (5) where *H* is 30, the number of years considered in this research.

$$Frequency_{i,j} = \frac{N_{i,j}}{\sum_{i=1}^{M} N_i} \quad (4)$$

$$\overline{Frequency_i} = \frac{\sum_{j=1}^{H} Frequency_{i,j}}{H} \quad (5)$$

The choice of research methods in a study is driven by the research questions it seeks to answer, and these questions are typically formulated in response to the research topic or problem being investigated. As a result, the diversity of research methods used can be thought of as the entropy of methods employed for different research topics in a given year. Equation (6) shows the variety formula of the *i*-th method in the *j*-th year. *T* is the number of research topics that appeared in the *j*-th year. The number of research topics is identified with the technique depicted in Section 3.3. $N_{j,k}$ is the number of all articles about the *k*-th research topic in the *j*-th year. $N_{i,j,k}$ is the number of articles about the *k*-th research topic and using the *i*-th method in the *j*-th year. We take the average application variety of the *i*-th method in all years as the overall variety of the *i*-th method, as shown in Equation (7) where *H* is 30, the number of years covered in this study.

$$Variety_{i,j} = -\sum_{k=1}^{T} \frac{N_{i,j,k}}{N_{j,k}} \log \frac{N_{i,j,k}}{N_{j,k}} \quad (6)$$

$$\overline{Variety_i} = \frac{\sum_{j=1}^{H} Variety_{i,j}}{H} \quad (7)$$

In this study, variety is used as a metric to quantify the breadth of research topics to which a given research method can be applied. A higher entropy value indicates that a research method can be utilized to study a wider range of research topics. For instance, if Research Method A has a higher entropy value than Research Method B, it means that Research Method A can be applied to a greater number of research topics compared to Research Method B.

## 4. Results

### 4.1 The automatic approach in this study

We answer RQ1 in this section. As stated at the beginning of this report, one outstanding feature of our study is its adoption of the automatic approach to addressing the first two research questions formulated for this investigation. The automatic classification of research methods was experimented with in two stages. The first stage involved an experiment with the extended dataset including noise while the second stage was an experiment on the cleaned dataset (see Section 3.2 and Figure 2). Table 3 presents the five-fold cross-validation results of the two-stage experiments with the micro $F_1$, macro $F_1$, and Samples $F_1$ as evaluation metrics. The formulas and explanations for the three metrics of evaluation can be found in Section 3.2 of this report.

| **Model** | **Expanded dataset** | | | **Cleaned dataset** | | |
|---|---|---|---|---|---|---|
| | **Micro-$F_1$** | **Macro-$F_1$** | **Samples-$F_1$** | **Micro-$F_1$** | **Macro-$F_1$** | **Samples-$F_1$** |
| BR-XGB | 0.618 | 0.649 | 0.562 | 0.750 | 0.758 | 0.675 |
| CC-XGB | 0.615 | 0.652 | 0.618 | 0.760 | 0.772 | 0.737 |
| LP-XGB | 0.564 | 0.597 | 0.595 | 0.736 | 0.756 | 0.660 |
| RAKEL-XGB | 0.604 | 0.633 | 0.541 | 0.753 | 0.762 | 0.680 |
| MLKNN | 0.318 | 0.297 | 0.258 | 0.369 | 0.282 | 0.314 |
| SCIBERT | 0.550 | 0.493 | 0.520 | 0.782 | 0.744 | 0.793 |
| SCIBERT-GRU-Att | 0.586 | 0.595 | 0.583 | 0.832 | 0.824 | 0.845 |
| SCIBERT-TextCNN | **0.632** | **0.664** | **0.628** | **0.850** | **0.848** | **0.859** |

**Table 3. Performance of eight different research method classification models**

Table 3 shows that SciBERT-TextCNN performs the best in the expanded, noisy dataset among all the classification models. We thus use SciBERT-TextCNN to complete confident learning. After cleaning the expanded dataset, we conducted experiments on the cleaned dataset and noticed that the classification accuracy of all classification models was improved. As SciBERT-TextCNN remains to yield the optimal classification results, we used it as the final model for the automatic classification of research methods on the unlabeled data.

Table 4 lists the five-fold cross-validation results of each method, using the classification quality measures Micro-P, Micro-R, and Micro-F1 (see Section 3.2). Among them, “Other methods” and “Content analysis” scored the lowest in Micro-F1. By further analyzing the data of these two methods, we find that the features of these two methods are more difficult to identify automatically than that of other research methods in this study. This is because “Other methods” actually contain multiple, different research methods (e.g., card sorting, information horizon). Similarly, “Content analysis” lacks a specific description of its features as a data collection technique.

| No. | Research method | Micro-P | Micro-R | Micro-$F_1$ |
| --- | --- | --- | --- | --- |
| 1 | Research diary/journal | 1.000 | 0.985 | 0.992 |
| 2 | Delphi study | 0.971 | 1.000 | 0.985 |
| 3 | Ethnography/field study | 0.979 | 0.971 | 0.974 |
| 4 | Focus groups | 0.987 | 0.933 | 0.959 |
| 5 | Bibliometrics | 0.939 | 0.939 | 0.938 |
| 6 | Historical method | 0.952 | 0.924 | 0.938 |
| 7 | Experiment | 0.938 | 0.919 | 0.928 |
| 8 | Transaction log analysis | 0.909 | 0.844 | 0.875 |
| 9 | Webometrics | 0.894 | 0.787 | 0.832 |
| 10 | Observation | 0.915 | 0.753 | 0.822 |
| 11 | Theoretical approach | 0.890 | 0.726 | 0.798 |
| 12 | Questionnaire | 0.857 | 0.763 | 0.797 |
| 13 | Interview | 0.847 | 0.642 | 0.720 |
| 14 | Think aloud protocol | 1.000 | 0.581 | 0.711 |
| 15 | Other | 0.526 | 0.909 | 0.657 |
| 16 | Content analysis | 0.618 | 0.672 | 0.641 |

**Table 4. Five-fold cross-validation results for each research method**

To further verify the classification results of our model, we tested the prediction results of the model on raw data. We randomly selected five samples for each method from the prediction results of unlabeled data and manually labeled them (N= 80) independently. We then calculated the Kappa value for the manually labeled results and the machine annotation results and obtained a Kappa value of 0.74 (McHugh, 2012). In manual coding or labeling, a Kappa value of 0.8 indicates an intercoder consistency that would be acceptable to most coders who participated in the coding process (Neuendorf, 2002). Although we only achieved the Kappa value of 0.74, it measures the coding or annotation consistency between machines and people in identifying and categorizing research methods. This is different from the intercoder consistency between or among human coders. In that sense, a Kappa value of 0.74 should be the acceptable outcome in this study, which proves that the automatic approach to classifying research methods can attain a similar accuracy of classification as the manual approach.

**4.2 Usage frequency and application variety: An overall view**

Table 5 displays the frequency of usage for each of the 16 research methods in our dataset. As shown, "Content analysis" is the most prevalent method in LIS research, with over one-sixth of the 26,965 articles analyzed reporting it as their research method. "Bibliometrics", "Experiment", "Theoretical approach", and "Questionnaire" are also frequently employed in LIS research, each accounting for around 10% of all the articles analyzed in this study.

| No. | Research method | Number of articles | Percent |
|---|---|---|---|
| 1 | Content analysis | 5,516 | 18.63 |
| 2 | Bibliometrics | 5,242 | 17.70 |
| 3 | Experiment | 4,590 | 15.50 |
| 4 | Theoretical approach | 3,855 | 13.02 |
| 5 | Questionnaire | 3,240 | 10.94 |
| 6 | Interview | 2,584 | 8.73 |
| 7 | Historical method | 1,085 | 3.66 |
| 8 | Webometrics | 952 | 3.22 |
| 9 | Transaction log analysis | 515 | 1.74 |
| 10 | Observation | 492 | 1.66 |
| 11 | Think aloud protocol | 459 | 1.55 |
| 12 | Other methods | 370 | 1.25 |
| 13 | Focus groups | 318 | 1.07 |
| 14 | Ethnography/field study | 293 | 0.99 |
| 15 | Research diary/journal | 80 | 0.27 |
| 16 | Delphi study | 19 | 0.06 |
| Total | | 29,610 | 100 |

**Table 5. Classification results of research methods**

In contrast, the remaining methods are used less frequently due to their unique characteristics. It is worth noting that the total number of articles listed in Table 5 is 29,610, which is greater than the initial 26,965 publications collected for this study. This is because some articles may report more than one research method as part of their research design.

Using the usage frequency of the 16 research methods examined in this study (as shown in Table 5) and the 18 research topics identified from our dataset (as presented in Table 4), we calculated the average usage frequency and average application variety scores for each of the 16 research methods (displayed in Table 6).

As per Equation (5), the average usage frequency value ranges from 0 to 1. The average usage frequency of a research method is the average annual usage frequency of that method from 1991 to 2021. For example, the average usage frequency of "Experiment" is 0.1547 in Table 6. This indicates that, on average, "Experiment" was used 15.47% of the time among all 16 methods per year over the 31-year period.

| No. | Method | $\overline{Frequency_i}$ | $\overline{Variety_i}$ |
|---|---|---|---|
| 1 | Content analysis | 0.1959 | 7.0840 |
| 2 | Bibliometrics | 0.1636 | 3.0589 |
| 3 | Experiment | 0.1547 | 3.6485 |
| 4 | Theoretical approach | 0.1470 | 5.4834 |
| 5 | Questionnaire | 0.1050 | 4.2097 |
| 6 | Interview | 0.0816 | 3.8934 |
| 7 | Historical method | 0.0397 | 2.2949 |
| 8 | Webometrics | 0.0301 | 2.1358 |
| 9 | Transaction log analysis | 0.0179 | 1.1965 |
| 10 | Observation | 0.0158 | 1.1350 |
| 11 | Think aloud protocol | 0.0150 | 1.1051 |
| 12 | Other methods | 0.0124 | 1.0152 |
| 13 | Focus groups | 0.0103 | 0.8462 |
| 14 | Ethnography/field study | 0.0090 | 0.7668 |
| 15 | Research diary/journal | 0.0032 | 0.2882 |
| 16 | Delphi study | 0.0014 | 0.1340 |

**Table 6. Overview of average usage frequency and application variety scores**

The application variety score can take a value greater than 1, as explained in Section 3.4. The entropy in this context measures the variety of research topics for which a research method can be used. However, the application variety value cannot be interpreted in the same way as usage frequency. An application variety value of zero (0) implies that one research method can only be applied to study one of the 18 research topics listed in

Table 6. When the application variety value of a research method increases beyond 0, it means that the method can be used to study more than one of the 18 research topics identified. The application variety value only evaluates the variety of research topics for which a research method can be employed, and the higher the value, the greater the variety of research topics for which the method can be used.

The mean value of the application variety for one research method represents the average variety of research topics for which that research method can be used over the 31 years covered in this study. For example, Table 6 shows that the mean application variety value for "Experiment" is 3.6485, which is much higher than that of the "Delphi study" (i.e., 0.1340). This implies that "Experiment" can be used to study more research topics than "Delphi study". Among the 16 research methods considered in this study, "Content analysis" has the highest average application variety score of 7.0840 (see Table 6). This indicates that "Content analysis" can be used to research a wide range of topics in LIS. It is partly for this reason that Chu and Ke (2017) refer to "Content analysis" as a super-method in the taxonomy of research methods they created based on applicability.

### 4.3 Usage frequency and application variety: A longitudinal view

In this section, we address RQ2 and examine the longitudinal usage frequency and application variety of research methods. Figure 5 illustrates the annual usage frequency of all 16 research methods considered in this study from 1991 to 2021.

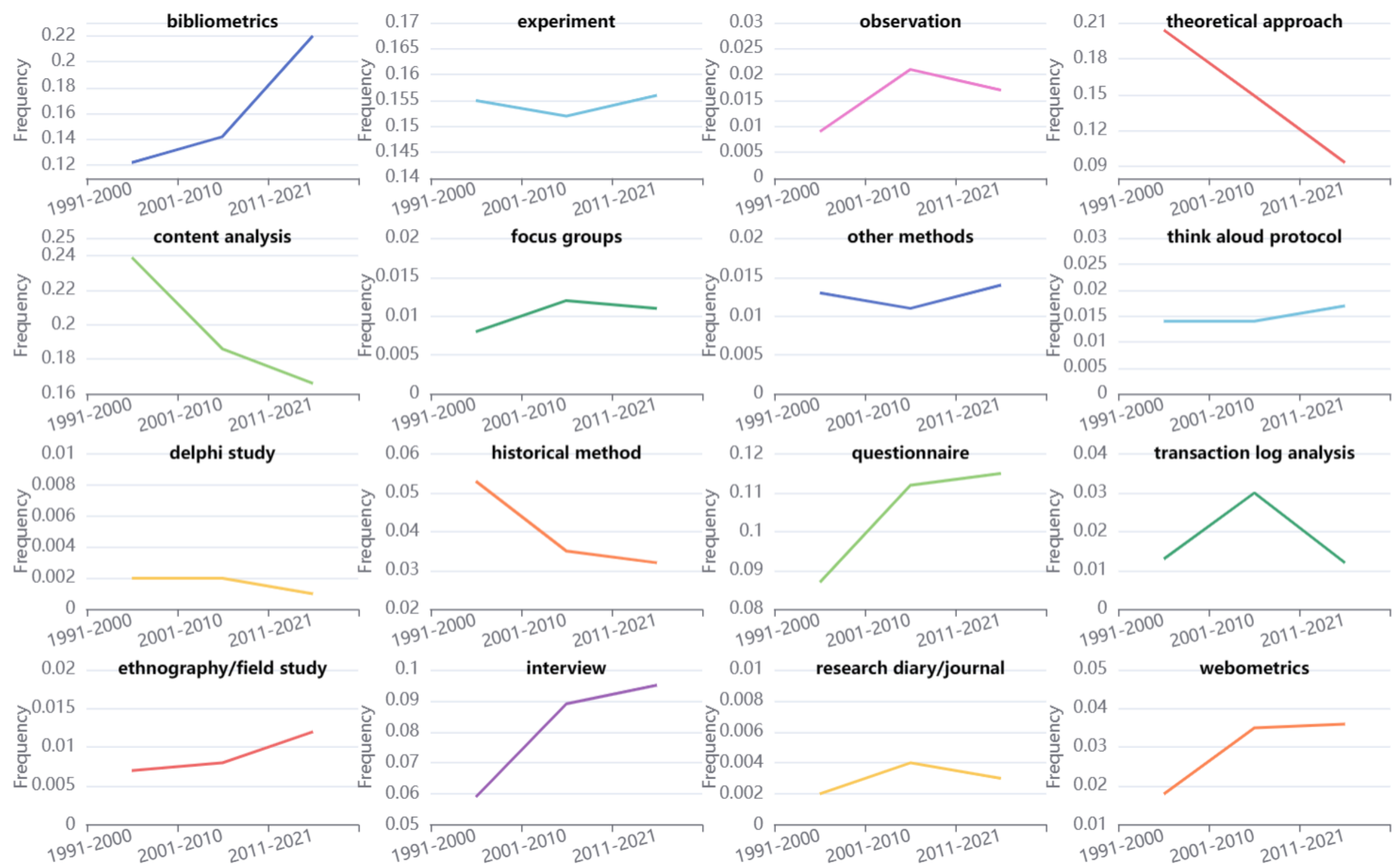

**Figure 5. Distribution of usage frequency of research methods from 1991 to 2021**

The figure demonstrates that the usage frequency of most research methods has remained stable over the 31-year period. However, the “Theoretical approach” and “Content analysis” exhibit a noticeable downward trend, while the “Interview” and “Questionnaire” show an upward trend. As the "Theoretical approach" is the primary method of conceptual research strategy, while “Interview”, “Questionnaire” and “Bibliometrics” are the primary methods of empirical research strategy, the increased usage frequency of the latter two methods is not surprising. This shift reflects the increasing adoption of the empirical research strategy by more LIS researchers in recent years. The usage frequencies of “Experiment”, “Observation”, “Focus group”, “Transaction log analysis” and “Research diary/journal” exhibit fluctuations over the 31-year period, as shown in Figure 5. Nevertheless, these methods remain major research methods in the LIS field in terms of usage frequency.

Figure 6 displays the annual application variety of the 16 research methods examined in this study from 1991 to 2021. Four different patterns in the application variety of research methods can be observed, as shown in Figure

6.

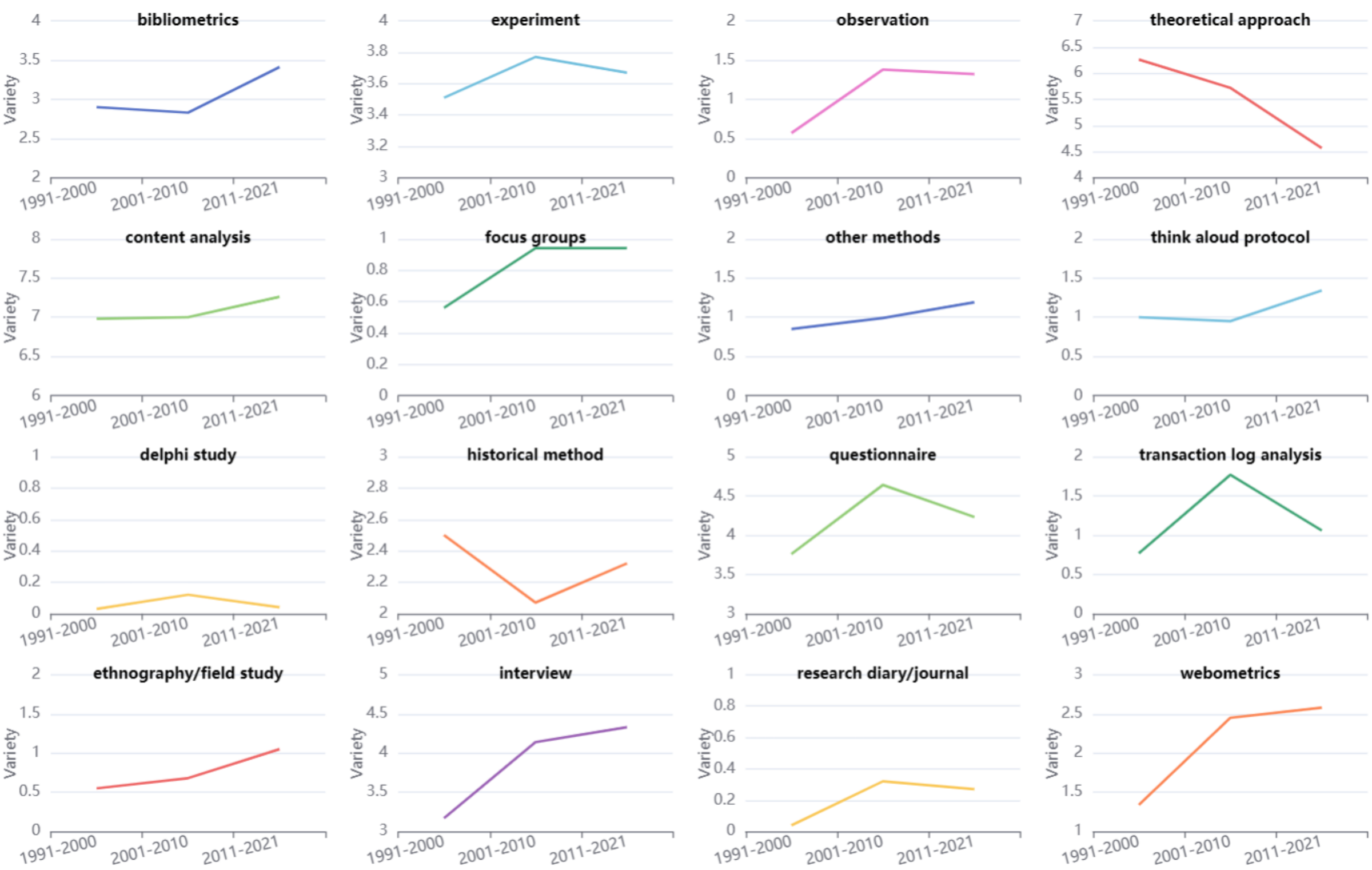

**Figure 6. Distribution of application variety of research methods from 1991 to 2021 in LIS**

The four patterns observed in Figure 6 represent four categories of research methods with different characteristics in LIS research (see Table 7). The first category, showing a pattern of overall increased application variety, encompasses two research methods: “Bibliometrics”, “Content analysis”, “Focus groups”, “Think aloud protocol”, “Ethnography/field study”, “Interview”, “Webometrics”. As the only research method originated from the LIS field, "Bibliometrics" continues to be applied to study a variety of research topics. The increased application variety of "Bibliometrics" reflects this method's wide applicability. The research methods that continue to increase in application variety also include “Focus groups”, “Think aloud protocol”, “Ethnography/field study”, “Interview”. As discussed earlier, LIS researchers have been paying more attention to the empirical research strategy in the 31 years we examined. The usage of “Focus groups”, “Think aloud protocol”, “Ethnography/field study”, “Interview” consequently increase not only in frequency but also in application variety.

“Experiment”, “Observation”, “Delphi study”, “Questionnaire”, “Transaction log analysis”, “Research diary/journal”, the six research methods in the second category in Table 7, share a pattern of increased first and then decreased application variety. While the “Questionnaire” used to be the dominating research method in LIS, its application becomes less widespread when scholars started to adopt other research methods for their investigations and pay more attention to the qualitative approach. This finding echoes what other researchers (e.g., Chu, 2015) have reported: the “Questionnaire” has been replaced by such research methods as “Content analysis” and “Experiment”. "Transaction log analysis", on the other hand, peaked its application variety when the internet and information retrieval (IR) systems (e.g., search engines and OPACs) entered the lives of the general public in the late 1990s and most of the time in the 2000s. A decent number of studies consequently adopted this research method to explore, for example, information seeking behavior and other research topics listed in Appendix C with transaction logs that were produced in abundance when the internet became the platform for numerous information systems, including IR systems.

| No. | Application Variety Pattern | Research Method |
|---|---|---|
| 1 | Overall increased application variety | Bibliometrics, Content analysis, Focus groups, Think aloud protocol, Ethnography/field study, Interview, Webometrics |
| 2 | Increased first and then decreased application variety | Experiment, Observation, Delphi study, Questionnaire, Transaction log analysis, Research diary/journal |
| 3 | Overall decreased application variety | Theoretical approach |
| 4 | Decreased first and then increased application variety | Historical method, |

**Table 7. Application variety pattern of research methods**

The third category consists of the “Theoretical approach”, which shows a continuous decline in application variety over the 31 years. The third category consists of the "Historical method". This reflects the shift in LIS research methods from conceptual to empirical research strategies.

### 4.4 Relationship between research topics and research methods

In this section, we aim to answer RQ3 by exploring changes in research topics in LIS publications from 1991 to 2021. We also intend to determine if any relationship exists between research topics and research methods or if certain research methods are often used for specific research topics. We identified 18 research topics using automatic means, as presented in Section 3.3, and present their percentage distribution in Table 8. Note that for discussion clarity, the topics in Table 8 are listed by percentage in descending order, while they are displayed alphabetically in Table 4.

| No. | Topic | Percentage |
|---|---|---|
| 1 | Information retrieval/models and algorithms | 8.37 |
| 2 | School libraries/Information literacy | 7.69 |
| 3 | Bibliometrics/Scientometrics | 7.34 |
| 4 | Information ethics/philosophy of information | 7.19 |
| 5 | Library history, history of L&I institutions | 7.09 |
| 6 | Organizational information activities | 6.72 |
| 7 | Open access publishing | 6.69 |
| 8 | Science evaluation | 5.79 |
| 9 | Information services | 5.44 |
| 10 | Information seeking behavior | 5.33 |
| 11 | Document delivery/Collections | 5.04 |
| 12 | Webometrics | 4.69 |
| 13 | Information organization/ Information management | 4.15 |
| 14 | Health information services/communities | 4.00 |
| 15 | Government policies/User education | 3.93 |
| 16 | Text processing/retrieval | 3.81 |
| 17 | Metadata/cataloguing | 3.42 |
| 18 | Digital information resources | 3.31 |

**Table 8. Percentage distribution of the 18 identified research topics**

Table 8 demonstrates that the first seven research topics, including "Information retrieval/models and algorithms," "School libraries/Information literacy," "Bibliometrics/Scientometrics," "Information ethics/philosophy of information," "Library history, history of L&I institutions," "Organizational information activities," and "Open access publishing," account for more than half of the total number of articles analyzed in this study. These research topics are the most significant ones in the LIS field from 1991 to 2021.

Figure 7 illustrates the changes in the 18 research topics over the 31-year period. Among them, "Information retrieval/models and algorithms", "Information organization/Information management", "Library history, history of L&I institutions", "Document delivery/Collections", and "Metadata/cataloguing" show a downward trend. This reflects the changing research focus of LIS scholars and the gradual maturity of some LIS research directions. "Information retrieval/models and algorithms" had represented a major research area in LIS for several decades before the start of the current study in 1990, thanks to the rapid development of information and database technologies. This research area has been gradually taken up by computer scientists in R&D, whereas LIS researchers mainly focus on application and evaluation.

In contrast, five research topics, namely "Open access publishing", "Health information services/communities", "Information services ", "Text processing/retrieval", and "Webometrics", show an upward trend over the 31 years. These research topics are all related to users, reflecting the user-centered approach LIS scholars take in selecting their research topics. Compared to the continuously declining "Information retrieval/models and algorithms," "Information services " is a relatively new research topic in LIS. Scholars before 1991 concentrated more on the system-centered approach than the user-centered one. As LIS researchers realize the importance of the user-centered approach, they conduct more research and publish more articles on "Information services."

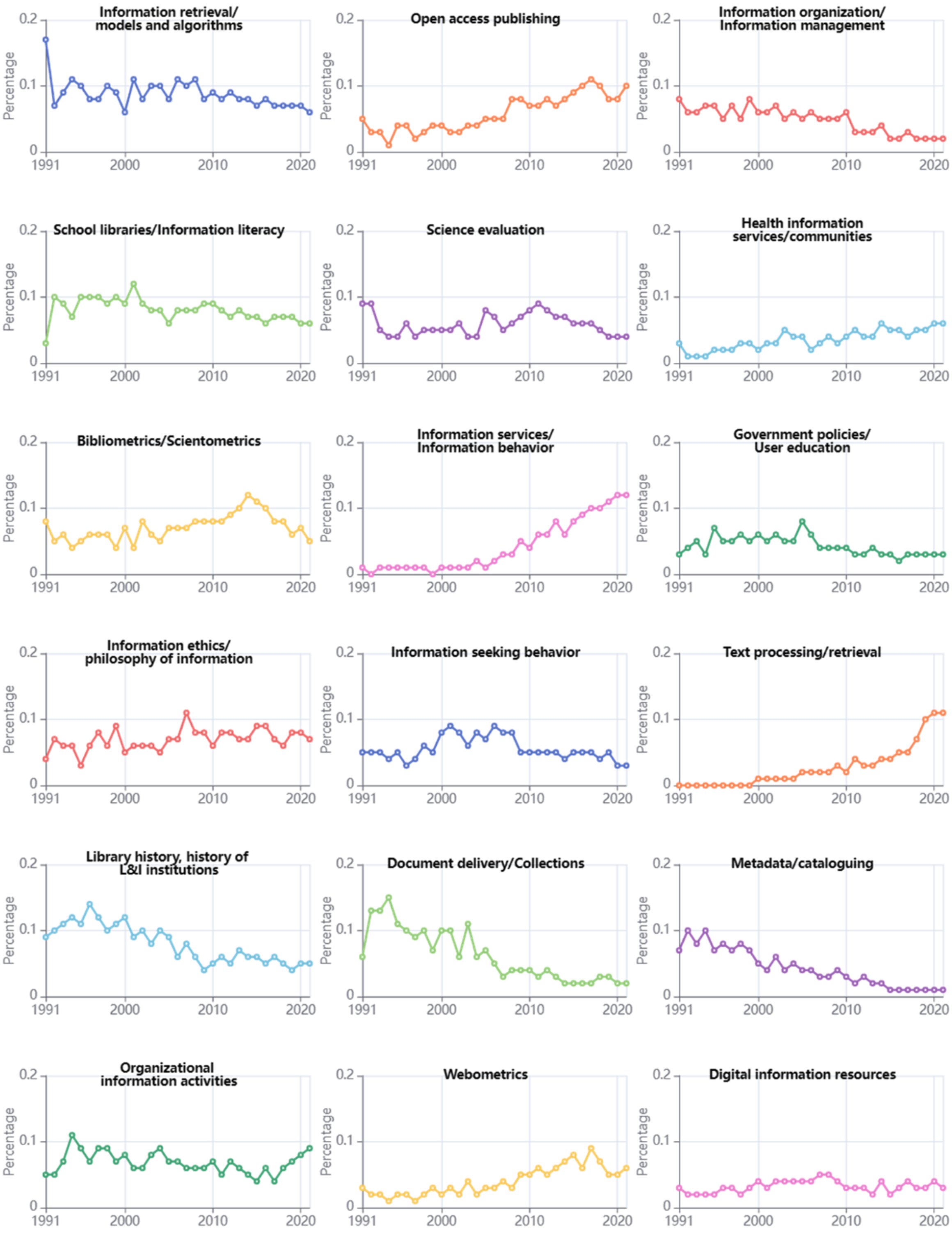

**Figure 7. Changes in the share of publications on different research topics**

Figure 7 also shows that publications on "School libraries/Information literacy", "Science evaluation", "Bibliometrics/Scientometrics", "Government policies/User education", "Information ethics/philosophy of information", "Information seeking behavior", "Organizational information activities", and "Digital information resources" remained relatively stable over the 31 years. This indicates that these research topics have been and may continue to be the focus of LIS scholars. The research topics explored from 1991 to 2021 changed, with an increasing emphasis on research projects related to users.

After presenting the longitudinal views for the 16 research methods and 18 research topics separately, it is time to determine the extent to which each research method was used for exploring individual research topics over the 31-year period. Figure 8 illustrates the findings, with darker colors indicating more frequent use of a research method. "Bibliometrics" is also moderately used for research topics on "Webometrics." "Bibliometrics" forms the basis for the "Webometrics" method and can be applied to examine the "Webometrics" topic. "Bibliometrics" is also chosen to address many specific aspects of library and information science, such as making acquisitions and journal subscription decisions.

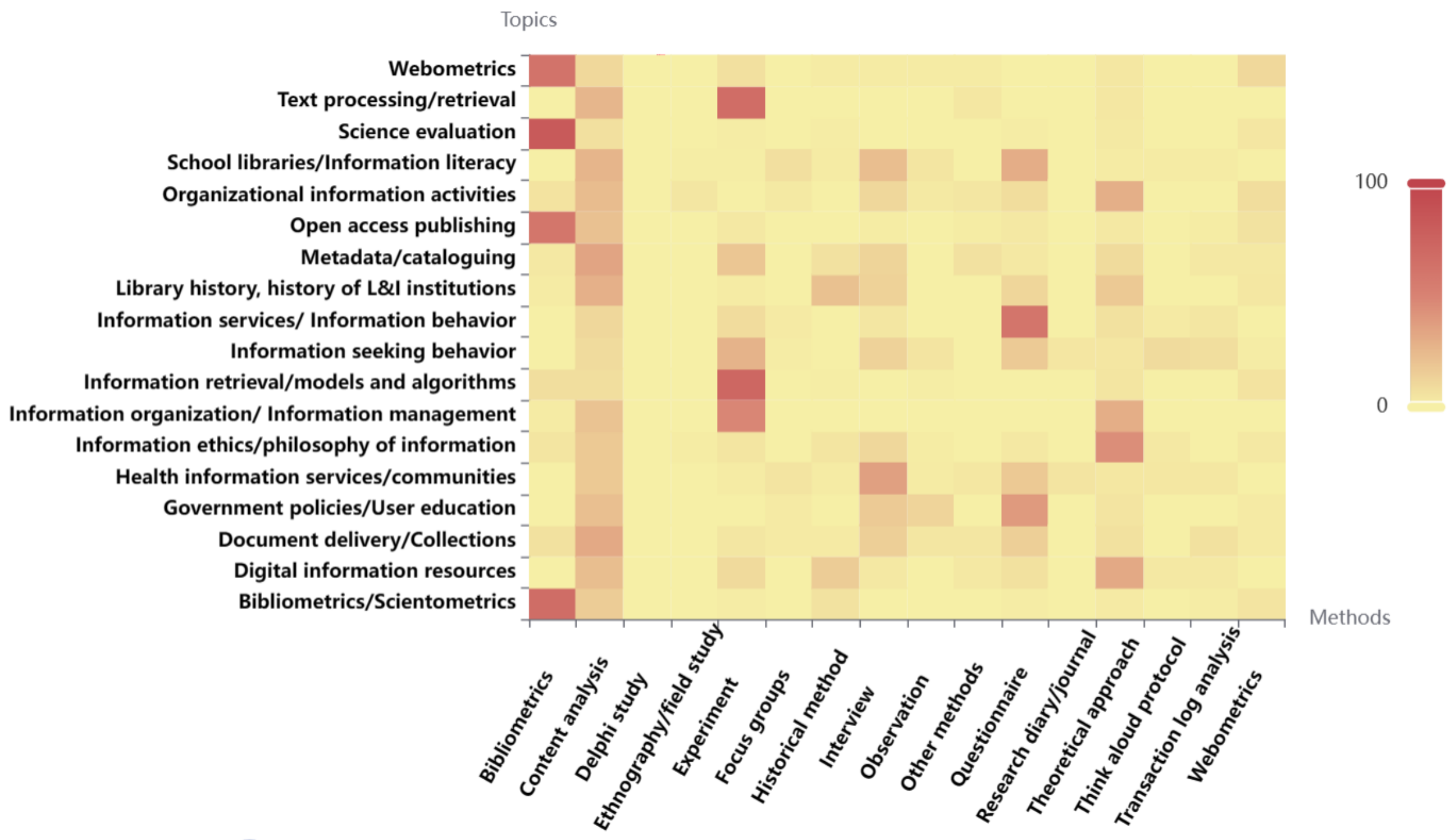

**Figure 8. Usage of Research Methods for LIS Research Topics in 1991-2021**

Furthermore, "Content analysis" has been utilized to investigate a wide range of research topics, although none of the applications seem to be extensive and inclusive. Many of the 18 research topics can be explored through "Content analysis," or "Content analysis" is required for investigating those topics. Similarly, "Interviews" are also frequently adopted to study most of the 18 research topics identified in this research. As a representative research method for collecting qualitative data, "Interviews" have been and will be increasingly used by LIS scholars to explore various research topics in the shift from the quantitative paradigm to the qualitative one.

Figure 8 shows that "Experiments" have been extensively used in research on "Text processing/retrieval," "Information retrieval/models and algorithms," "Information seeking behavior," and "Information organization/Information management." Among these four topics, "Experiments" are the most appropriate method for research on "Text processing/retrieval" and "Information retrieval/models and algorithms" because researchers can only test out various techniques and algorithms for text processing and information retrieval through experiments. While experiments on techniques and algorithms do not involve human participants, research on "Information seeking behavior" commonly requires the use of "Experiments" to determine the type of "Information seeking behavior" exhibited by human participants in the process. Additionally, the "Theoretical approach" appears to be another research method frequently applied in studies on "Information ethics/Philosophy of information." The research topic of "Information ethics/philosophy of information," especially the latter part, requires conceptual analysis and theoretical discussion, which is exactly what the "Theoretical approach" can offer as a method.

As shown, Figure 8 and the related discussion reveal the overall connection between research topics and research methods from 1991 to 2021. To provide a better view of this relationship, we created an interactive diagram at https://chengzhizhang.github.io/research/research_methods/research_topic_method.html. The interactive diagram enables us to observe the changes in research topics and the corresponding research methods used each year. One can use the Pause button to the left of the Year-axis at the bottom of the interactive map to view the relationship between research topics and research methods in the LIS field for any particular year from 1991 to 2021. This interactive map truly portrays the dynamic relationship by year between research topics and research methods in LIS over 31 years, a unique presentation that can be viewed both yearly or longitudinally.

## 5. Discussion

### 5.1 Unique contributions to studies on research methods

Our fundamental purpose of this study is to provide a panoramic view of research methods reported in LIS publications from the perspective of usage frequency and application variety through the machine learning approach. In addition, the present study makes two unique contributions to investigations on research methods. First, we develop and adopt a machine learning approach to achieving the objective of automatic classification of research methods, automatic identification of research topics as well as exploration of the relationship between research methods and research topics. This approach effectively overcomes the limitation encountered by related studies in the past relying on manual content analysis, namely, on average only several dozen (e.g., Yontar & Yalvac, 2000) or several hundred (e.g., Tuomaala, Järvelin & Vakkari, 2014) articles can be examined in a study. By comparison, the current study uses a dataset of 26, 000 articles thanks to the machine learning approach in this study takes. Second, we not only explore the usage frequency of research methods, but also investigate the application variety of research methods.

The results we obtain in this study using the machine learning approach are similar to that done via manually in terms of research method classification. For example, Chu (2015) examined articles from the same three journals as this study, finding that “Theoretical approach”, ”Experiment”, and “Content analysis” were among the top used research methods, each accounting for 21.7%, 17.8% and 19% respectively. The same three research methods are also among the top chosen methods in this study. Ma and Lund (2021) reported similar findings as the current study although they worked with articles from a different set of journals. Yet, both Chu (2015) and Ma and Lund (2021) analyzed a much smaller dataset. Our deep learning techniques enable us to conduct this line of research using a larger dataset and over a longer period of consecutive years. Studies relying on manual analysis cannot observe the continuous change of what is under study in a long period. It is thus difficult to provide a panorama of research methods in the LIS field.

Both Tuomaala, Järvelin and Vakkari (2014) and Ma and Lund (2021) looked at research methods and research topics in LIS in their studies. However, they basically treated research methods and research topics as two separate, unrelated themes and simply examined and reported each independent of the other. In contrast, the present study explores the usage frequency of research methods along and also in conjunction with research topics, namely, what research methods are used for studying which research topic or application variety in short. Therefore, the current study presents a fuller picture the utilization and application of research methods in LIS from1990 to 2021.

**5.2 An innovative methodology with practical implication**

The automatic identification technology of research methods in LIS literature we have developed represents the-state-of-art R&D efforts compared with related research. For instance, Eckle-Kohler, Nghiem and Gurevych (2013) used machine learning techniques while Zhang, Tam and Cox (2021) adopted simple rule-based algorithms in their respective research.

Our methodology improves the traditional approach to studying research methods applied in LIS investigations. However, we must point out that our methodology is not a complete replacement of the manual method. Rather it is a complement to the manual approach. The research methods of a discipline are changing and evolving, so must be the coding scheme for categorizing research methods. A well-rounded coding scheme can only be developed by human beings instead of machines. In that sense, the manual approach to future studies in this line of research is indispensable. The machine learning approach or other related automated techniques would no doubt help expand the scope of similar investigations with better efficiency.

The machine learning developed and adopted for the present research can of course be applied to disciplines other than library and information science. Our plan is to apply the same methodology to articles from LIS journals beyond the three we examine in the current study. We also intend to try our methodology with datasets obtained from disciplines other than library and information science. We tentatively decide to start with the domain of information systems in consideration of the studies that are already done using the traditional approach (e.g., Avison, Dwivedi, Fitzgerald & Powell, 2008; Palvia, Pinjani & Sibley, 2007).

### 5.3 Limitations of this study

We have used a variety of machine learning techniques to perform automatic classification of research methods and automatic identification of research topics. However, our methodology does not perform well with some research methods such as “Content analysis”, and we plan to improve it in our future research. We used the coding scheme of research methods developed by Chu and Ke (2017), which is comprehensive but perhaps not detailed enough. Some methods (e.g., “Experiment”) could be further divided in order to better categorize research methods reported in publications. We intend to revise that coding scheme in the future when conducting additional research in this area.

## 6. Conclusions

The results presented and discussed above allow for several conclusions to be drawn regarding the three research questions posed in this study.

Firstly, there has been a shift from the conceptual research strategy to the empirical research strategy in the research topics explored by LIS scholars from 1991 to 2021. This indicates that LIS researchers are increasingly using data collection techniques that enable them to gather qualitative data (such as interviews), rather than focusing mainly on quantitative data collection methods (such as questionnaires). This paradigm shift reflects the choices of research methods by LIS scholars for their inquiries.

Secondly, the research topics explored by LIS scholars over the three decades considered in this study have shifted from system-centered issues (such as "Information retrieval/models and algorithms") to research projects with an emphasis on user-centered subjects (such as "Information services "). LIS researchers are no longer limiting their studies to the systems in libraries and information centers. Instead, they are paying more attention to users in their research because the user dimension ultimately determines the quality of information services offered by libraries and information centers.

Third, the 16 research methods examined in this study differ in their application variety, with "Content analysis" having the highest mean application variety score (7.084) and "Delphi study" the lowest (0.134). This suggests that "Content analysis" is a versatile research method that can be used to study a wide range of topics, while "Delphi study" is limited in its applicability. Different research topics require different research methods, and the suitability of a particular research method for a specific topic depends on its features. Therefore, the choice of research method should always be guided by the research topic under study.

Fourth, the relationships between the 18 research topics identified in this study and the 16 research methods commonly used in LIS are both revealing and dynamic over the 31-year period studied. For instance, “Bibliometrics” is heavily used for researching “Science evaluation”, while “Content analysis” is applied to study most of the 18 research topics, although none of the applications seems extensive and inclusive. The interactive map created in this study provides a unique view of the dynamic relationships between research topics and methods over time.

Fifth, this study adopted a machine learning approach to automatically identify research topics and classify research methods in the dataset. Unlike previous studies that rely on manual classification, our experiment results demonstrate that the automatic classification model works well, achieving an accuracy in classifying research methods that is identical to the manual approach. This outcome is a promising development for researchers working on text classification in general and the categorization of research methods in particular.

**Acknowledgment**

This study has received support from the National Natural Science Foundation of China (Grant No. 72074113) and the Major Project of the National Social Science Fund of China (Grant No. 20&ZD332, 20ZDA039).

**Appendix A**

Coding schema for research methods in Chu and Ke (2017).

| **Method** | **Definition** |
|---|---|
| Bibliometrics (including citation analysis, informetrics, & scientometrics) | Bibliometrics is a method used for collecting publication and citation data. |
| Content analysis (including discourse analysis, & secondary analysis) | Content analysis refers to collecting data by conducting systematic examination of texts or other passages in the contexts of their use. |
| Delphi study | The Delphi method is generally used for collecting data with a questionnaire from a group of experts to address a research problem in order to reach consensus and make forecasts via several rounds of exchanges. |
| Ethnography/field study | Ethnography and field study share many characteristics in data collection. Both can be applied when collecting data using multiple techniques, such as observation and interview, in a natural setting where participants live or work. |
| Experiment | Experiment is an established method for collecting data by following a procedure to test what is studied in either a laboratory or field setting, corresponding to laboratory experiments and field experiments described in |
| | Palvia et al.'s (2007) list of research methods. |
| Focus groups | As a research method, focus groups refer to data collection via discussion of a research problem between a moderator and a group of participants. |
| Historical method | Historical method refers to collecting data by examining, synthesizing, summarizing, and interpreting existing published and unpublished materials related to a historical research problem. |
| Interview | Interview is a data collection technique where individual participants are asked questions relating to a research problem. |
| Observation | Observation is a method for gathering data via carefully and attentively watching and making notes on the subject being studied. |
| Questionnaire (including index, inventory, scale, & test) | Questionnaire, often known as survey, is a technique for data collection using a predefined list of questions. |
| Research diary/journal | Research diary or journal is a technique used to gather data about events, activities, thoughts, reflections, or other aspects by an individual who keeps the diary over a period of time. |
| Theoretical approach (e.g., conceptual analysis, modeling, & theory building) | Theoretical approach, as a research method, is a technique for gathering data through conceptual analysis, theoretical examination, or similar activities. |
| Think aloud protocol | Think aloud protocol is a research method intended to collect data about participants' cognitive activities via the verbal reports of their thoughts, called think alouds, while taking part in an experiment or performing some task. |
| Transaction log analysis | Transaction log analysis, as a research method, gains momentum when computerized systems are used for information processing and access. |
| Webometrics (including link analysis, cybermetrics, & altmetrics) | Webometrics is defined as bibliometrics in the web environment, where webpages and websites are generally regarded as publications; with inlinks (i.e., links a webpage or site receives) being considered as citations and outlinks (i.e., links a webpage or site makes to others) being considered as references. |
| Other methods (e.g., action research, card sorting, information horizon) | Research methods other than the 15 mentioned above. |

**Appendix B**

**The top 10 most frequent keywords of each low-frequency research method**

| **Method** | **Top 10 Keywords** | **Ratio of valid documents to the total number of search return results.** |
|---|---|---|
| Delphi study | study, Delphi, expert, knowledge, identify, lead, process, design, theory, management | 64.0% |
| Ethnography/field study | social, approach, data, finding, practice, ethnographic, participant, understand, analysis, user | 57.5% |
| Focus groups | focus, group, data, library, interview, user, health, provide, support, participant | 49.5% |
| Historical method | historical, paper, study, library, thousand, article, development, analysis, history, change | 59.0% |
| Observation | user, observation, library, search, interview, finding, result, system, participant, behavior | 80.0% |
| Research diary/journal | diary, user, participant, social, search, people, interaction, relationship, process, partner | 66.0% |
| Think aloud | participant, data, design, task, process, interview, system, library, finding, web | 50.5% |
| Transaction log analysis | log, web, query, result, system, data, library, session, engine, term | 64.0% |
| Webometrics | web, study, site, link, result, measure, search, user, university, citation | 56.0% |
| Other methods | user, search, participant, eye, system, design, attention, card, task, time | 54.0% |

**Appendix C**

**18 identified research topics in LIS.**

| **No.** | **Topic** | **Definitions** |
|---|---|---|
| 1 | Scientific and professional communication | Studies related to scholarly communication subjects, including citation patterns, cooperation and other related subjects. |
| 2 | Digital information resources | Studies related to digital information resource construction (digital library, digital archive management, digital museum) and digital information resource services. |
| 3 | Document delivery/Collections | Studies related to document delivery and academic library research collections. |
| 4 | Rural libraries/User education | Studies related to rural library, information literacy education. |
| 5 | Health information services/communities | Studies related to health information systems, health information seeking and health information communities. |
| 6 | Information ethics/philosophy of information | Studies related to philosophy of information and information ethics. |
| 7 | Information organization/ Information management | Studies related to information organization and information management. |
| 8 | Information retrieval/models and algorithms | Studies related to models, algorithms of information retrieval. |
| 9 | Information seeking behavior | Studies related to information seeking behavior. |
| 10 | Information services | Studies related to information service quality assessment, service satisfaction assessment, improvement of information service quality, and |

| | |
|---|---|
| | other related subjects |
| 11 Library history, history of L&I institutions | Studies related to the history of libraries, librarianship, LIS, and other related subjects. |
| 12 Metadata/cataloguing | Studies related to metadata and cataloguing in information retrieval. |
| 13 Open access publishing | Studies related to database, resource, and scientific impact of open access. |
| 14 Organizational information activities | Studies related to organizational information activities, including theory, management, security of organizational information, and other subjects related to organizational information activities. |
| 15 School libraries/Information literacy | Studies related to school libraries and information literacy education for students. |
| 16 Science evaluation | Studies related to design, implementation science evaluation. |
| 17 Text processing/retrieval | Studies related to processing and retrieval of text. |
| 18 Webometrics | Studies related to the construction and use of information resources, structures and technologies on the Web drawing on bibliometric and informetric approaches. |